\documentclass[12pt]{article}

\catcode`@=11

\renewcommand{\@seccntformat}[1]{{\csname the#1\endcsname.\quad}}
\usepackage{theorem}

\newtheorem{theorem}{Theorem}[section]

\newtheorem{proposition}[theorem]{Proposition}

{\theorembodyfont{\rmfamily}
\newtheorem{definition}[theorem]{Definition}
\newtheorem{remark}{Remark}

}

\newcommand{\beginproof}{\medskip\noindent{\bf Proof.~}}
\newcommand{\beginproofof}[1]{\medskip\noindent{\bf Proof of #1.~}}
\newcommand{\beginproofdotless}{\medskip\noindent{\bf Proof}}
\newcommand{\finishproof}{\hspace{0.2ex}\rule{1ex}{1ex}}

\newcounter{tenumerate}

\newlength{\mathindent}
\catcode`@=12

\newcommand{\refeq}[1]{{\rm (\ref{#1})}}
\newlength{\longeqmarginwidth}
\newlength{\longeqwidth}
\newlength{\longeqskiplength}
\newcommand{\longeq}[1]{
\settowidth{\longeqmarginwidth}{{\Large \{~\}}~(\theequation)}
\setlength{\longeqwidth}{\textwidth}
\addtolength{\longeqwidth}{-\longeqmarginwidth}
\left. \parbox{\longeqwidth}{\begin{eqnarray*} #1
\end{eqnarray*}} \right\}}

\def\gr
{\mathrel{\kern0.1em\hbox{\vbox{\baselineskip =-100pt\lineskip=.2ex%
\halign{##\cr\vbox{\hrule height .35pt width 1em} \cr
\kern .33em \hbox{$ \scriptscriptstyle  \circ$} \cr
\vbox{\hrule height .35pt width 1em} \cr}}\kern-0.22em}}}

\newcommand{\Vzero}[2]{\mathord{\stackrel{\textstyle\kern1pt\circ}
{\smash V\vbox to6pt{}}}\vphantom{V}_{#1}^{#2}}

\newcommand{\function}[2]{:#1 \longrightarrow #2}

\newcommand{\set}[2]{\left\{\hspace{0.2ex} #1 \left|\: #2
\right. \right\}}

\newcommand{\df}{\stackrel{\rm def}{=}}

\newcommand{\tr }{{\rm tr}}

\newcommand{\scr}{\mathcal}

\newcounter{operator}

\usepackage{amsfonts}
\usepackage{amsmath}
\title{An Upper Bound on the Threshold Quantum Decoherence Rate}
\author{Alexander A. Razborov \thanks{Institute for Advanced Study,
Princeton, US, on leave from Steklov Mathematical Institute, Moscow,
Russia, razborov@ias.edu. Supported by the State of New Jersey,
The Bell Companies Fellowship, The James D. Wolfensohn Fund, and
The Ellentuck Fund.}}
\begin{document}
\maketitle

\begin{abstract}
Let $\eta_0$ be the supremum of those $\eta$ for which every poly-size
quantum circuit can be simulated by another poly-size quantum circuit with
gates of fan-in $\leq 2$ that tolerates random noise independently
occurring on all wires at the constant rate $\eta$. Recent fundamental
results showing the principal fact $\eta_0>0$ give estimates like
$\eta_0\geq 10^{-6}\mbox{--}10^{-4}$, whereas the only upper bound known
before is $\eta_0\leq 0.74$.

In this note we improve the latter bound to $\eta_0\leq 1/2$, under the
assumption ${\bf QP}\not\subseteq {\bf QNC^1}$. More generally, we show
that if the decoherence rate $\eta$ is greater than 1/2, then we can not
even store a single qubit for more than logarithmic time. Our bound also
generalizes to the simulating circuits allowing gates of any (constant)
fan-in $k$, in which case we have $\eta_0\leq 1-\frac 1k$.
\end{abstract}

\section{Introduction}
Whereas it is still too premature to rush for any definite conclusions
about the prospects of practical quantum computing, some important issues
presumably confronting any implementation scheme have been identified. And
it seems to be of more or less universal agreement that the most serious
of these issues is the problem of decoherence due to the interaction of
the quantum system with its environment. For this reason it has become an
extremely important task for theoreticians to identify the type and amount
of noise against which arbitrary quantum computations can be protected
with at most polynomial slow-down.

The noise model most often considered in the literature (and the one we
are sticking to in the current note) is that of {\em local noise} (see
e.g. \cite{Aha}). In this model, errors  occur on every
wire in the quantum circuit independently with a certain probability
$\eta$. They occur even on those wires which are currently not
involved in the computation but simply transmit information from one
level to another. As a compensation for this (otherwise devastating)
assumption, the circuit is also allowed to have a parallel
architecture and perform operations on disjoint qubits
simultaneously. The basic question is how small should the noise rate
$\eta$ be so that every poly-time quantum computation can be
transformed into another poly-time quantum computation resistant to
this kind of noise.

\medskip
The early (and relatively easy) lower bound on the threshold value
$\eta_0$ below which the noise-resistant computation becomes possible was
given in \cite{BeV}. Namely, they showed $\eta_0\geq\Omega(1/n)$, where
$n$ is the number of gates in the protected circuit. \cite{Shor2} improved
this to $\eta_0\geq (\log n)^{-O(1)}$. Finally, \cite{AhB,Kit} proved the
fundamental result that $\eta_0$ is separated from 0 by some universal
constant not depending on the circuit size at all, and various estimates
of the quality of these error-correction schemes give lower bounds on
$\eta_0$ in the range $10^{-6}\mbox{--}10^{-4}$ (we have listed above only
some of the most central work in this direction; further references can be
found e.g. in the monographs \cite{NiC,KVS} and in the expositions
\cite{Aha,Got,Pre}).

Much less work has been done on {\em lower} bounds for fault-tolerant
quantum computations (that is, {\em upper} bounds on the threshold
$\eta_0$). \cite{AhB2} showed that $\eta_0\leq 0.96$, as long as the
simulating circuit is allowed to have gates of fan-in $\leq 2$
only. Then it was established in the beautiful paper \cite{ABI*} that
fault-tolerant computation is impossible for {\em any} constant rate
of noise $\eta$ without an uninterrupted supply of fresh qubits
initialized in (known) pure states. Finally, using ideas quite
different from those in \cite{AhB2,ABI*}, \cite{HaN} resently showed
that $\eta_0\leq 0.74$ (they were also able to prove $\eta_0\leq 0.5$,
but in a much more restricted model of the adversarial noise not
necessarily independent on wires, and only for certain specific 
unitary quantum gates on two qubits).

\smallskip
As any normal (or, depending on the point of view, arrogant) complexity
theorist, the author is rather wary about the activity of determining the
exact value of various quantities previously established to be an absolute
constant. Still, the range $\eta_0\in [10^{-6}\mbox{--}10^{-4},\ 0.74]$ is
rather embarrassing and humiliating. Also, it is conceivable that this is
the order of magnitude of $\eta_0$ that will have a final say on whether
quantum computers will some day become available. For these reasons, the
author believes that this case should be granted an exception, and that
the questions like $\eta_0\stackrel ?\geq 0.1$ or $\eta_0\stackrel ?\geq
0.01$ are of great {\em theoretical} value.

\bigskip
In this note we take the next modest step toward narrowing the gap
and show that $\eta_0\leq 1/2$, unless ${\bf QP}= {\bf QNC^1}$. For
doing that we prove, in the style of \cite{ABI*}, that when the noise
rate $\eta$ is greater than 1/2, any quantum computation completely
collapses within logarithmic time. Our proof is much simpler than the
proofs in \cite{AhB2,HaN}, and it also has a straightforward
generalization to quantum circuits allowing gates of arbitrary fan-in
$k$, with the corresponding bound $\eta_0\leq 1-\frac 1k$.

\section{Preliminaries and the main result}

In this note we exclusively deal with mixed state quantum circuits
\cite{AKN,Kit}. For this reason we completely skip the usual quantum
formalism pertinent to the standard (unitary) model of quantum computing
(for that see e.g. \cite{NiC,KVS}) and immediately proceed to
density matrices.

\bigskip
For a finite dimensional Hilbert space $\scr N$, ${\bf L}(\scr N)$ is the
set of linear operators on $\scr N$. ${\bf D}(\scr N)\subseteq {\bf
L}(\scr N)$ is the set of {\em density matrices on $\scr N$}: $\rho\in
{\bf D}(\scr N)$ if and only if $\rho$ is Hermitian, positive
semi-definite and $\tr(\rho) =1$. ${\bf D}(\scr N)$ is a convex subset in
${\bf L}(\scr N)$.

A linear mapping $T\function{{\bf L}(\scr N)}{{\bf L}(\scr M)}$ is a
{\em quantum operation} iff it is trace-preserving and completely
positive (that is, sends positive semi-definite operators to positive
semi-definite operators and retains this property even after taking a
tensor product with the identity operator on an arbitrary Hilbert
space). The set of all quantum operations $T\function{{\bf L}(\scr
N)}{{\bf L}(\scr M)}$ is denoted by ${\bf T}(\scr N,\scr M)$. Quantum
operations take density matrices to density matrices, and the tensor
product of quantum operations is a quantum operation.

For $\eta\in [0,1]$, {\em depolarization at the rate $\eta$} is the
specific quantum operation $\scr E_\eta\in {\bf T}(\scr N,\scr N)$
given by
\begin{equation}\label{noise}
\scr E_\eta(\rho)\df (1-\eta)\rho + \frac{\eta\tr(\rho)}{\dim (\scr
N)}I_{\scr N},
\end{equation}
where by $I_{\scr N}$ we denote the identity operator on $\scr
N$. Physically, it corresponds to the process in which the system
described by the density matrix $\rho$ gets depolarized (= replaced
with the completely mixed state $I_{\scr N}/\dim(\scr N)$) with
probability $\eta$, and is left untouched otherwise.

\begin{remark}
Depolarization is exactly the noise model considered in
\cite{ABI*,HaN}. \cite{AhB2} used a slightly different model
(``dephasing'', or measurement in a given basis), but it is
straightforward to see that their method actually applies to the
depolarizing noise as well.
\end{remark}

The {\em trace distance} $D(\rho,\sigma)$ between two density matrices
$\rho$ and $\sigma$ on the same space $\scr N$ is defined as
$D(\rho,\sigma)\df \frac 12||\rho-\sigma||_{\tr}$, where
$||\cdot||_\tr$ is the trace norm. It is equal to the maximal
difference in the results of measuring $\rho$ and $\sigma$ in the same
basis one can achieve.

\begin{sloppypar}
\begin{proposition}\label{convex}
For any $p_1,\ldots,p_n\geq 0$ with $p_1+\cdots+p_n=1$ and
$\rho_1,\ldots,\rho_n,\sigma_1,\ldots,\sigma_n\in {\bf D}(\scr N)$,
$D(\sum_ip_i\rho_i, \sum_ip_i\sigma_i)\leq \sum_ip_iD(\rho_i,\sigma_i)$.
\end{proposition}
\end{sloppypar}

\begin{proposition}[\protect{\cite[Theorem 9.2]{NiC}}] \label{monotone}
For any $\rho,\sigma\in {\bf D}(\scr N)$ and $T\in {\bf T}(\scr N,\scr
M)$, $D(T(\rho),T(\sigma))\leq D(\rho,\sigma)$.
\end{proposition}

\smallskip
Quantum circuits with mixed states \cite{AKN,Kit} are just
ordinary parallel circuits allowing gates from a prescribed set $\scr
G$ of quantum operations. More formally, let $\scr B$ be a
two-dimensional Hilbert space (``qubit''), and let $\scr B^{\otimes
n}\df \scr B\otimes\cdots\otimes\scr B$ ($n$ times). For any partition
$[n]= I_1\stackrel .\cup\ldots\stackrel .\cup I_w$ of the ground set
$[n]\df\{1,\ldots, n\}$, $\scr B^{\otimes n}$ can be naturally
decomposed as $\scr B^{\otimes |I_1|}\otimes\cdots\otimes \scr
B^{\otimes |I_w|}$, and we will freely use this notation. A {\em
quantum gate} is a quantum operation $T\in {\bf T}(\scr B^{\otimes k},
\scr B^{\otimes\ell})$, $k$ being called {\em fan-in} of the gate
$T$. Let in particular $\scr G_k$ be the set of all gates with fan-in
$\leq k$.

For a family $\scr G$ of quantum gates, a {\em quantum circuit $Q$
over $\scr G$} is any sequence of quantum operations

\begin{sloppypar}
\begin{equation}\label{circuit}
\scr B^{\otimes n_0}\stackrel{T_0}{\longrightarrow} \scr B^{\otimes
n_1} \stackrel{T_1}{\longrightarrow} \ldots
\stackrel{T_{t-1}}{\longrightarrow} \scr B^{\otimes n_t},
\end{equation}
in which every $T_i$ allows a decomposition of the form
\begin{equation}\label{decomposition}
\longeq{[n_i] &=& K_{i1}\stackrel .\cup K_{i2} \stackrel .\cup \ldots
\stackrel .\cup K_{iw_i};\\
\protect{[n_{i+1}]} &=& L_{i1}\stackrel .\cup L_{i2}
\stackrel .\cup \ldots \stackrel .\cup L_{iw_i};\\
T_i &=& \bigotimes_{\nu=1}^{w_i}T_{i\nu},\ T_{i\nu}\in {\bf T}(\scr
B^{\otimes |K_{i\nu}|}, \scr B^{\otimes |L_{i\nu}|})\cap \scr G.}
\end{equation}
The {\em depth} $Depth(Q)$ of the circuit \refeq{circuit} is $t$, and its
{\em width} $Width(Q)$ is $\max\{n_0,n_1,\ldots,n_t\}$. For an input
$\rho\in {\bf D}(\scr B^{\otimes n_0})$, the {\em final state of running
$Q$ on $\rho$} is simply
$$
Q(\rho)\df T_{t-1}T_{t-2}\ldots T_1T_0(\rho).
$$
For $\eta\in [0,1]$, the {\em perturbed circuit} $Q_\eta$ is obtained by
interlacing the computational steps $T_0,T_1,\ldots, T_{t-1}$ with the
noise operators $\scr E_\eta$ independently depolarizing all qubits with
probability $\eta$. That is,
\begin{equation}\label{noise_circuit}
Q_\eta(\rho)\df T_{t-1}\scr E_\eta^{\otimes n_{t-1}}T_{t-2}\scr
E_\eta^{\otimes n_{t-2}}\ldots \scr E_\eta^{\otimes n_2}T_1\scr
E_\eta^{\otimes n_1}T_0(\rho).
\end{equation}
\end{sloppypar}

\begin{remark}
Introducing fresh qubits into the system (unavoidable due to the result of
\cite{ABI*}) can be thought of as a special quantum operation from ${\bf
T}(\scr C,\scr B)$ of fan-in 0 (cf. \cite{Kit}), and does not require any
special treatment in our framework.
\end{remark}

\smallskip
\cite{ABI*} called the quantum circuit $Q$ of the form \refeq{circuit}
{\em worthless} if for every $\rho\in {\bf D}(\scr B^{\otimes n_0})$,
$D(Q(\rho),\ 2^{-n_t}\cdot I_{\scr B^{\otimes n_t}})\leq 1/100.$ Following
their suit, we introduce the following (slightly more relaxed) definition:
\begin{definition}
A quantum circuit $Q$ is {\em practically worthless} if for every
$\rho,\sigma\in {\bf D}(\scr B^{\otimes n_0})$, $D(Q(\rho),
Q(\sigma))\leq 1/100$.
\end{definition}
Thus, whereas worthless circuits do not produce any output at all,
practically worthless circuits can compute only constant Boolean
functions 0 and 1.

Our main result can be now stated as follows:
\begin{theorem}\label{main}
For every constants $k>0$ and $\eta>1-\frac 1k$ there exists $C>0$
such that the following holds. For every quantum circuit $Q$ over
$\scr G_k$ with $Depth(Q)\geq C\log Width(Q)$, the perturbed circuit
$Q_\eta$ is practically worthless.
\end{theorem}

This theorem says that as long as the noise rate exceeds $1-\frac 1k$,
every two initial states become totally indistinguishable within
$O(\log n)$ steps, where $n$ is the number of qubits allowed in the
system. In particular, no Boolean function $f\in {\bf QP}\setminus
{\bf QNC^1}$ can be computed fault-tolerantly. Also, it will become
clear from the proof that the same conclusion holds if $n$ is
understood as the number of qubits {\em participating in the final
measurement}. If, for example, the circuit $Q_{\eta}$ is required to
output the result by measuring only one qubit in the final state (as
is often the case in the standard unitary model), then it becomes
practically worthless already within a {\em constant} number of
steps.

\section{The proof}

Before we start, let us remark that the proof can be re-casted in the
completely combinatorial style of \cite{AhB2}. Namely, the same recursion
as the one used below can also show that if $\eta>1-\frac 1k$ then with
high probability the computational graph of the circuit becomes
disconnected after deleting from it faulty wires. We, however, prefer more
analytical version of the proof (that was actually found first, and is
inspired in part by \cite{ABI*}) as it appears to us more suggestive of
potential improvements and generalizations.

\bigskip
Let $Q$ be a quantum circuit of the form \refeq{circuit}, and $\eta\in
[0,1]$. For $\rho\in {\bf D}(\scr B^{\otimes n_0})$, denote by
$\rho_i$ the density matrix obtained at the $i$th level of the faulty
computation \refeq{noise_circuit} before applying the noise operator
$\scr E_\eta^{\otimes n_i}$. That is,
$$
\rho_i\df T_{i-1}\scr E_\eta^{\otimes n_{i-1}}T_{i-2}\scr
E_\eta^{\otimes n_{i-2}}\ldots \scr E_\eta^{\otimes n_2}T_1\scr
E_\eta^{\otimes n_1}T_0(\rho).
$$
Note for the record that $\rho_0=\rho$ and $\rho_t=Q_\eta(\rho)$.

Next, for $A\subseteq [n_i]$, let $\rho_i|_A$ be the result of tracing
out the density matrix $\rho_i$ with respect to all qubits in
$[n_i]\setminus A$ (cf. \cite{ABI*}). Then the effect of applying the
noise operator $\scr E_\eta^{\otimes n_i}$ on $\rho_i$, and, more
generally, on all its reduced submatrices $\rho_i|_B$ can be expressed
as follows (see \cite{ABI*}):
\begin{equation} \label{noise_action}
\scr E_\eta^{\otimes n_i}(\rho_i|_B) = \sum_{A\subseteq
B}\eta^{|B|-|A|}(1-\eta)^{|A|}(\rho_i|_A \otimes (I_{\scr B
}/2)^{\otimes(|B|-|A|)}),
\end{equation}
where the notation $\rho_i|_A \otimes (I_{\scr B}/2)^{\otimes(|B|-|A|)}$
corresponds to the partition of $B$ into $A$ and $B\setminus A$.

Let now $\sigma\in {\bf D}(\scr B^{\otimes n_0})$ be another initial
density matrix. For $i=0,\ldots,t$ and $n\geq 0$ let
$$
d_{in}(\rho,\sigma)\df\max_{A\subseteq [n_i]\atop |A|\leq
n}D(\rho_i|_A,\sigma_i|_A).
$$
Then, clearly,
\begin{equation} \label{rec_one}
d_{00}(\rho,\sigma)=0,\,\  d_{0n}(\rho,\sigma)\leq 1\ (n\geq 1).
\end{equation}

In order to get a recursive bound on $d_{in}$, we additionally assume that
the fan-in of all gates in the circuit $Q$ is bounded by $k$. Fix
$A\subseteq [n_{i+1}]$ with $|A|\leq n$. Recalling the decomposition
\refeq{decomposition}, let $\Gamma\df\set{\nu\in [w_i]}{L_{i\nu}\cap
A\neq\emptyset}$ be the set of all quantum gates involved in the
computation of qubits from $A$, and $B\df\bigcup_{\nu\in\Gamma}K_{i\nu}$
be the set of all their inputs. Then $|B|\leq k\cdot|\Gamma|\leq kn$, and
by Proposition \ref{monotone} (applied to the quantum operation
$\bigotimes_{\nu\in\Gamma}T_{i\nu}$ and density matrices $\scr
E_\eta^{\otimes n_i}(\rho_i|_B), \scr E_\eta^{\otimes n_i}(\sigma_i|_B)$),
\begin{eqnarray*}
D(\rho_{i+1}|_A, \sigma_{i+1}|_A) &\leq&
D(\rho_{i+1}|_{\bigcup_{\nu\in\Gamma}L_{i\nu}},
\sigma_{i+1}|_{\bigcup_{\nu\in\Gamma}L_{i\nu}})\\ &\leq& D(\scr
E_\eta^{\otimes n_i}(\rho_i|_B), \scr E_\eta^{\otimes n_i}(\sigma_i|_B)).
\end{eqnarray*}
Furthermore, applying Proposition \ref{convex} to the representation
\refeq{noise_action},
\begin{eqnarray*}
D(\scr E_\eta^{\otimes n_i}(\rho_i|_B), \scr
E_\eta^{\otimes n_i}(\sigma_i|_B) &\leq& \sum_{A\subseteq
B}\eta^{|B|-|A|}(1-\eta)^{|A|}D(\rho_i|_A,\sigma_i|_A)\\&\leq&
\sum_{m=0}^{|B|} {|B|\choose
m}\eta^{|B|-m}(1-\eta)^md_{im}(\rho,\sigma)\\&\leq& \sum_{m=0}^{kn} {kn\choose
m}\eta^{kn-m}(1-\eta)^md_{im}(\rho,\sigma),
\end{eqnarray*}
where for the last inequality we additionally used the fact that, by
definition, $d_{im}(\rho,\sigma)$ is monotone in $m$. Combining the
last two inequalities, and recalling that $A\subseteq [n_{i+1}]$ with
the property $|A|\leq n$ was chosen arbitrarily, we get
\begin{equation}\label{rec_two}
d_{i+1,n}(\rho,\sigma)\leq \sum_{m=0}^{kn} {kn\choose
m}\eta^{kn-m}(1-\eta)^md_{im}(\rho,\sigma).
\end{equation}

It is now straightforward to see that the recursion \refeq{rec_one},
\refeq{rec_two} has the exact solution
$$
d_{in}(\rho,\sigma)\leq 1-f_i^n,
$$
where the coefficients $f_i\in [0,1]$ are given by
$$
f_0\df 0,\ \ f_{i+1}\df (\eta+(1-\eta)f_i)^k
$$
(and $0^0\df 1$). Applying the inequality $(1-x)^k\geq 1-kx\ (0\leq x\leq
1)$ with $x:=(1-\eta)(1-f_i)$, we get from here $f_{i+1}\geq
1-\theta(1-f_i)$, where $\theta\df k(1-\eta)<1$ is a constant. This gives
us $f_i\geq 1-\exp(-\Omega(i))$. In particular, if $t\geq C\log n_t$ for a
sufficiently large constant $C>0$ then $f_t\geq 1-\frac 1{n_t^2}$ and
$d_{t,n_t}(\rho,\sigma)\leq O(1/n_t)$.

We have shown that $D(Q_\eta(\rho), Q_\eta(\sigma))\leq O(1/n_t)$ for
every pair $\rho,\sigma\in {\bf D}(\scr B^{\otimes n_0})$, which
completes the proof of Theorem \ref{main}.

\section*{Acknowledgment}

I am grateful to Julia Kempe for several stimulating and useful
discussions, and to Scott Aaronson for pointing out to me the
reference \cite{HaN}. 


\end{document}